\documentclass[12pt]{article}
\usepackage{graphicx}
\usepackage{psfig,epsfig}
\usepackage{caption}

\def\lsim{\raise0.3ex\hbox{$\;<$\kern-0.75em\raise-1.1ex\hbox{$\sim\;$}}}
\def\gsim{\raise0.3ex\hbox{$\;>$\kern-0.75em\raise-1.1ex\hbox{$\sim\;$}}}
\def\ib#1#2#3{           {\it ibid. }{\bf #1} (19#2) #3}
 
\def\np#1#2#3{           {\it Nucl. Phys. }{\bf #1} (19#2) #3}
\def\pl#1#2#3{           {\it Phys. Lett. }{\bf #1} (19#2) #3}
\def\pr#1#2#3{           {\it Phys. Rev. }{\bf #1} (19#2) #3}

\def\prl#1#2#3{          {\it Phys. Rev. Lett. }{\bf #1} (19#2) #3}

\def\sjnp#1#2#3{         {\it Sov. J. Nucl. Phys. }{\bf #1} (19#2) #3}
\def\jetp#1#2#3{         {\it Sov. Phys. JETP }{\bf #1} (19#2) #3}

\def\ppnp#1#2#3{           {\it Prog. Part. Nucl. Phys. }{\bf #1} (19#2) #3}

\def\eq#1{{eq.~(\ref{#1})}}
\def\Eq#1{{Eq.~(\ref{#1})}}
\newcommand{\ea}{\end{array}}

\def\Frac#1#2{\frac{\displaystyle{#1}}{\displaystyle{#2}}}
\def\lsim{\raise0.3ex\hbox{$\;<$\kern-0.75em\raise-1.1ex\hbox{$\sim\;$}}}
\def\gsim{\raise0.3ex\hbox{$\;>$\kern-0.75em\raise-1.1ex\hbox{$\sim\;$}}}
\begin{document}
\thispagestyle{empty}
\begin{titlepage}
\begin{center}
\hfill hep-ph/9905405\\
\hfill FTUV/99-21\\
\hfill IFIC/99-22\\
\hfill SISSA 31/99/EP\\
\hfill{December 15, 1999}\\
\vskip 0.3cm
{\bf \Large A potential test of the CP properties and Majorana nature
of neutrinos}
\end{center}
\normalsize
\vskip1cm
\begin{center}
\baselineskip=13pt
{\bf S. Pastor~$^a$}\footnote{ E-mail: pastor@sissa.it},
{\bf J. Segura~$^b$}\footnote{ E-mail: javi.segura@umh.es},
{\bf V.B. Semikoz~$^c$}\footnote{ E-mail: semikoz@izmiran.rssi.ru}
{\bf and J.W.F. Valle~$^d$}\footnote{ E-mail: valle@flamenco.ific.uv.es}\\
\end{center}
\begin{center}
\baselineskip=13pt
{\sl $^a$ SISSA--ISAS and INFN, Sezione di Trieste}\\
\baselineskip=12pt
{\sl Via Beirut 2-4,I-34013 Trieste, ITALY}\\
\vglue 0.8cm
\end{center}
\begin{center}
\baselineskip=13pt
{\sl $^b$ Instituto de Bioingenier\'{\i}a \& \\
Departamento de Estad\'{\i}stica
y Matem\'atica Aplicada\\
Universidad Miguel Hern\'andez, Edificio La Galia}\\
\baselineskip=12pt
{\sl 03206 Elche, Alicante, SPAIN}\\
\vglue 0.8cm
\end{center}
\begin{center}
\baselineskip=13pt
{\sl $^c$ Institute of Terrestrial Magnetism, the Ionosphere and
Radio\\ Wave Propagation of the Russian Academy of Sciences, IZMIRAN}\\
\baselineskip=12pt
{\sl Troitsk, Moscow region, 142092, RUSSIA}\\
\vglue 0.8cm
\end{center}
\begin{center}
\baselineskip=13pt
{\sl $^d$ Instituto de F\'{\i}sica Corpuscular - C.S.I.C.\\
Departament de F\'{\i}sica Te\`orica, Universitat de Val\`encia\\}
\baselineskip=12pt
{\sl 46100 Burjassot, Val\`encia, SPAIN}\\
\baselineskip=12pt
{\sl http://neutrinos.uv.es}\\
\end{center}

\begin{abstract}
The scattering of solar neutrinos on electrons may reveal their CP
properties, which are particularly sensitive to their Majorana
nature. The cross section is sensitive to the neutrino dipole moments
through an interference of electro-magnetic and weak amplitudes.  We
show how future solar neutrino experiments with good angular
resolution and low energy threshold, such as Hellaz, can be sensitive
to the resulting azimuthal asymmetries in event number, and could
therefore provide valuable information on the CP properties and the
nature of the neutrinos, provided the solar magnetic field direction
is fixed.
\end{abstract}
\end{titlepage}
\vskip 1cm

\section{Introduction}

It has long been realized that, on general grounds, gauge theories
generally predict that, if neutrinos are massive, they should be
Majorana particles, unless protecting symmetries are imposed or arise
accidentally \cite{cpmaj}. Even though lepton-number-violating
processes such as neutrinoless double beta decay are intrinsically
related to the Majorana nature of neutrinos in a gauge theory
\cite{betabeta}, their search as so far yielded only negative results~
\cite{hdmo}.  It has also been shown in the early 80's that gauge
theories with Majorana neutrinos contain additional CP violating
phases without analogue in the quark sector~\cite{cpmaj}. Although
these are genuine physical parameters of the theory, as they show up
in $\Delta L =2$ neutrino oscillations \cite{cpmajexp}, their effects
are also suppressed by the smallness of neutrino masses, relative to
the typical neutrino energies available at accelerator and even
reactor experiments.

Among the non-standard properties of neutrinos the electro-magnetic
dipole moments \cite{BFD,Shrock82,Nieves82,MoPal,KIM} play an
important conceptual role, since they can potentially signal the
Majorana nature of neutrinos.  Neutrino transition electro-magnetic
moments \cite{BFD} are especially interesting because their effects
can be resonantly enhanced in matter~\cite{RFSP} and provide an
attractive solution of the solar neutrino problem \cite{akhmedov97}
without running in conflict with astrophysics~\cite{Raffelt}.

For pure left-handed neutrinos the weak interaction amplitude on
electrons does not interfere with that of the electro-magnetic
interaction, since the weak interaction preserves neutrino helicity
while the electro-magnetic does not. As a result the cross section
depends quadratically on the neutrino electro-magnetic form factors.
However if there exists a process capable of converting part of the
initially fully polarized neutrinos, then an {\sl interference term}
arises proportional to the neutrino electro-magnetic form factors, as
pointed out e.g. in ref. \cite{Barbieri}.  This term depends on the
angle between the component of the neutrino spin transverse to its
momentum and the momentum of the outgoing recoil electron.  Therefore
the number of events measured in an experiment exhibits an {\sl
asymmetry} with respect to the above defined angle.  The asymmetry
will not show up in terrestrial experiments even with stronger
magnetic fields, since only in the Sun the neutrino depolarization
would be resonant and only in the solar convective zone one will find
a magnetic field extended over such a region (about a third of the
solar radius wide). At earth-bound laboratory experiments the
helicity-flip could be caused only by the presence of a neutrino mass
and is therefore small \cite{grimus}, in a way analogous to the case
of neutrino $\Delta L =2$ lepton-number-violating neutrino
oscillations~\cite{cpmajexp}. Exotic couplings to scalars might change
this feature, but there are relatively strong limits.
In contrast for a relatively modest large-scale solar magnetic field
in the convective region $B_{\perp} \sim 10^4$ G and a neutrino
magnetic moment of the order $10^{-11} \mu_B$, where $\mu_B$ is the
Bohr magneton, one has $\mu_{\nu} B_{\perp} L \sim 1 $ since $L \sim
L_{conv} \simeq 2 \times 10^{10}$ cm is the width of the convective
zone. Such a spin-flip process may depolarize the solar neutrino flux
at a level where neutrino electro-magnetic properties may reveal the
Majorana nature of neutrinos (or alternatively, the solar magnetic
field structure).

In this paper we show that the resonant enhancement of neutrino
conversions induced by Majorana transition moments can provide
valuable hints on the true nature of neutrinos and their CP
properties, in a way which is not suppressed by the small neutrino
mass.  Our proposed test requires the careful investigation of
neutrino-electron scattering for neutrinos from the Sun at future
solar neutrino experiments with good angular resolution and low energy
threshold. One such proposed experiment is Hellaz \cite{ Hellaz}. For
completeness and pedagogy we also include a discussion of the
Dirac-type magnetic moment or electric dipole moment~\cite{VVO}.

\section{Neutrino Electro-Magnetic Properties }

The most general effective interaction Lagrangian describing the
electro-magnetic properties of Majorana neutrinos has been first given
in ref.~\cite{BFD} in terms of the fundamental two-component spinors.
The connection with conventional four-component description can be
found in ref.~\cite{cpmaj}. Other equivalent presentations are given
in \cite{MoPal,KIM} and the corresponding matrix element between
one-particle neutrino states for a real ($q^2=0$) photon, can be
written as~\cite{Raffelt}
\begin{equation}
< p',s',j | {\cal L}_{eff} | p,s,i > =
\bar{u}^{s'}_j (p') \Gamma^{ij}_\lambda (p,p') u^s_i (p) A^\lambda (q)=
\bar{u}^{s'}_j (p') i \sigma_{\lambda \rho} q^\rho
(\mu_{ij} + id_{ij} \gamma_5) u^s_i (p) A^\lambda (q)
\label{4}
\end{equation}
where $i,j$ denote the mass labels of the neutrinos, the indices $s$
and $s'$ specify helicities, while the $u$'s are the standard wave
functions of the Dirac equation and $q=p-p'$. Here $\mu_{ij}$ and
$d_{ij}$ are the {\em magnetic} and {\em electric dipole moments},
respectively. From the hermiticity condition for the Lagrangian one
can relate
\cite{Shrock82,Nieves82} the form factors of the $i \to j$ process and
its inverse,
\begin{equation}
\mu_{ij} = \mu^*_{ji}  \qquad d_{ij} = d^*_{ji}
\label{8}
\end{equation}
Note that in the diagonal $i=j$ case, both $\mu$ and $d$ must be {\sl
real} according to \eq{8}. For further constraints on the form factors
$\mu_{ij}$ and $d_{ij}$ one must assume something about the neutrino
nature and/or invariance under the CP symmetry~\cite{BFD,MoPal}.
These interactions arise only from loops in a gauge theory like the
Standard Model and are therefore calculable from first principles.
However, in most gauge theories magnetic moments are expected to be
small. For discussions see ref.~\cite{MoPal}.

Majorana neutrinos can have only off-diagonal ($i \neq j$) form
factors, called transition moments~\cite{BFD}, while if the neutrinos
are Dirac particles, just as the charged leptons, both diagonal and
off-diagonal moments can exist.  Let us assume that the effective
Lagrangian is invariant under a CP transformation, ${\cal L}_{eff} =
CP{\cal L}_{eff}(CP)^{-1}$.  A Dirac field transforms under CP as $CP
\Psi_i (\vec{x},t) (CP)^{-1}=\eta_i C \Psi^*_i (-\vec{x},t)$, where 
$\eta_i$ is a phase factor and $C$ is the charge conjugation matrix
($C^{-1} = C^\dagger = C^T = -C$) \cite{Nieves82,MoPal,KIM}.  If we
apply this to the $\sigma_{\mu \nu}$-part of the effective Lagrangian,
the CP-transformed $i \to j$ part will contribute to the $j \to i$ process
and vice versa. The result for $\Gamma_\lambda$ implies that
\begin{equation}
\bar{u}_j (p') \Gamma^{ij}_\lambda (q) u_i (p) =
\eta^*_i \eta_j \bar{u}_j (p') \bar{\Gamma}^{ij}_\lambda (q) u_i (p)
\label{10}
\end{equation}
where $\bar{\Gamma}^{ij}_\lambda$ is equal to $\Gamma^{ij}_\lambda$
with the change $\gamma_5 \to -\gamma_5$. \Eq{10} implies that
the form factors obey the relations
\begin{equation}
\frac{\mu_{ij}^*}{\mu_{ij}} = -\frac{d_{ij}^*}{d_{ij}} = \eta_i \eta^*_j
\label{11}
\end{equation}

A Majorana neutrino is its own anti-particle. It is easy to check in
this case that in \eq{4} both the $ij$ and the $ji$ terms in the
Lagrangian will contribute to the $ij$ form factors. One finds that
for {\em mass} eigenstates \cite{Nieves82}
\begin{equation}
< p',j | {\cal L}_{eff} | p,i > =
\bar{u}_j (p') i \sigma_{\lambda \rho} q^\rho
[(\mu_{ij}-\mu_{ji})+ i(d_{ij}-d_{ji}) \gamma_5]u_i (p) A^\lambda
\label{15}
\end{equation}
Finally, from the hermiticity condition \eq{8} one gets
$\mu_{ij}-\mu_{ji} = 2i \mbox{Im}(\mu_{ij})$ and $d_{ij}-d_{ji} = 2i
\mbox{Im}(d_{ij})$.  Therefore we conclude that a Majorana neutrino
has no diagonal electro-magnetic factors and that the transition form
factors $\mu_{ij}$ and $d_{ij}$ are both {\sl pure imaginary},
irrespective of whether or not one assumes CP invariance \cite{BFD}.
Thus Majorana neutrinos can only possess {\sl transition} magnetic or
electric dipole moments. Let us now check whether CP invariance
restricts them.  If CP is conserved, a Majorana neutrino is a CP {\em
eigenstate}, with a phase $\eta_{CP} = \pm i$ \cite{BFD}.  Considering
the invariance of ${\cal L}$ for the Majorana case under CP one gets a
condition similar to \eq{10} and \eq{11}. There are two physically
interesting cases to consider: two neutrino species involved in \eq{4}
can be either both active, weakly interacting neutrinos, or one of
them can be sterile. Moreover, for each of these cases, there are {\sl
two} possible CP-conserving cases, depending on the relative CP sign
of the neutrinos involved
\begin{enumerate}
\item Case ($+-$): $(\eta_i,\eta_j)=(\pm i, \mp i)$, then 
$\mu_{ij}$ survives and $d_{ij}=0$.  This is a {\sl pure magnetic}
transition, and includes the Dirac-type magnetic moment if one of the
neutrinos is sterile.
\item Case ($++$): $(\eta_i,\eta_j)=(\pm i, \pm i)$, then 
$\mu_{ij}=0$ and $d_{ij}$ survives~\cite{BFD}.
This is a {\sl pure electric} transition.
\end{enumerate}

On the contrary, as emphasized by Wolfenstein~\cite{BFD}, if CP is not
conserved both magnetic and electric dipole moments will contribute to
the neutrino-electron scattering cross section.  The general
properties of Dirac and Majorana neutrino electro-magnetic dipole
moments are summarized in table~\ref{majorana}.
\begin{table}
\centerline{
\begin{tabular}{|c|c|c|}
\hline
Case & Hermiticity & Hermiticity + CP \\
\hline
Dirac $i=j$ & $\mu_{ii}$ and $d_{ii}$ real & $d_{ii}=0$ \\
\hline
Dirac $i \neq j$ & 
$\mu_{ij}= \mu^*_{ji}$ and $d_{ij}= d^*_{ji}$ &
$\mu_{ij}$ and $id_{ij}$ relatively real \\
\hline
Majorana $i=j$ & $\mu_{ii}=d_{ii}=0$ & --- \\
\hline
& \multicolumn{1}{c|}{$\mu_{ij}$ and $d_{ij}$}
& \multicolumn{1}{c|}{Case ($+-$): $d_{ij}=0$} \\
\cline{3-3} \raisebox{3ex}[0cm][0cm]{Majorana $i \neq j$}
& pure imaginary
& Case ($++$): $\mu_{ij}=0$\\
\hline
\end{tabular}
}
\caption{General properties of neutrino electro-magnetic 
dipole moments}
\label{majorana}
\end{table}

Note that the above discussion is completely general and covers all
types of Majorana transition moments, active-active and
active-sterile.  In particular it covers the active-sterile case with
zero mass splitting (Dirac diagonal case). In what follows we will
focus mainly on active-active Majorana transition moments, as well as
the Dirac diagonal case.

\section{Dipole moments for flavor states}

We have discussed so far the restrictions upon the neutrino
electro-magnetic dipole moments for mass eigenstates. Since we are
interested in possible interference terms between weak and
electro-magnetic interactions in neutrino-electron scattering, we
shall present the corresponding matrix elements in terms of flavor
states.  For simplicity we will restrict ourselves to the
two-generation case, and for definiteness the $\nu_e-\nu_a$ pair,
where $\nu_a$ can be an active neutrino (for instance $\nu_\mu$) or a
sterile neutrino $\nu_s$.  In this case the mixing matrix contains a
CP violating phase for the case of Majorana
neutrinos~\cite{cpmaj,cpmajexp}. Such phase is absent if the two
neutrinos are Dirac type, since in this case it can be removed by
field redefinition, as expected in analogy with the quark sector,
where CP violation sets in only for three generations.

The $\bar{u}\Gamma_\lambda u$ matrix element can be written as
in~\eq{4} but for flavor eigenstates $\nu_{e,a}$, with
$\mu_{\nu_e \nu_a}$ and $d_{\nu_e \nu_a}$.  Here the restrictions on
$\mu$ and $d$ for Dirac mass states still apply, and in particular for
the diagonal case both $\mu_{\nu_e}$ and $d_{\nu_e}$ are real.

The CP-violating phase which is present in the mixing matrix of a
theory with two Majorana neutrinos may be introduced as the
$e^{i\beta}$ phase in the $2\times 2$ mixing matrix,
\begin{equation}
\left (\matrix{\nu_e \cr \nu_a}\right ) =
\left (\matrix{c & se^{i\beta} \cr -s & ce^{i\beta}}\right ) 
\left (\matrix{\nu_1 \cr \nu_2}\right ) =
\left (\matrix{c & s \cr -s & c}\right ) 
\left (\matrix{1 & 0 \cr 0 & e^{i\beta}}\right ) 
\left (\matrix{\nu_1 \cr \nu_2}\right )
\label{23}
\end{equation}
Here $c\equiv \cos \theta, s\equiv \sin \theta$, where $\theta$
denotes the leptonic mixing angle. Let us define now in
\eq{15} the real parameters $\mu'_{ij} \equiv$ Im$(\mu_{ij})$ and
$d'_{ij}\equiv$ Im$(d_{ij})$. The expression for \eq{15} is then as
follows
\begin{equation}
\bar{u}_j (p') i \sigma_{\lambda \rho} q^\rho 2i
(\mu'_{ij} + i d'_{ij} \gamma_5) u_i (p)
\label{24}
\end{equation}
If we introduce now the weak states according to \eq{23}, one gets two
contributions to the $\nu_e-\nu_a$ amplitude, corresponding to
$i=1,j=2$ and $i=2,j=1$.  Using the hermiticity condition one has
$\mu'_{21}=-\mu'_{12}$ and $d'_{21}=-d'_{12}$, and one can define the
electro-magnetic dipole moments for {\em flavor states} as follows
($\kappa = \mu, d$ and note that $\kappa_{ae} = \kappa^*_{ea}$)
$$
\kappa_{ea} \equiv \kappa_{\nu_e \nu_a} = 
2i (c^2 e^{i\beta} + s^2 e^{-i\beta}) \kappa'_{12} =
2i (\cos \beta + i(c^2 - s^2) \sin \beta) \kappa'_{12}
$$
\begin{equation}
\kappa_{ae} \equiv \kappa_{\nu_a \nu_e} = 
2i (c^2 e^{-i\beta} + s^2 e^{i\beta}) \kappa'_{21} =
2i (\cos \beta - i(c^2 - s^2) \sin \beta) \kappa'_{21}
\label{28}
\end{equation}

We conclude then that a pair of Majorana neutrinos (weak states) has,
in general, {\em complex} dipole moments. This is a consequence of the
CP phase from the mixing matrix.  The particular CP-conserving cases
correspond to the values $\beta = 0, \pi/2$. Therefore, when assuming
CP invariance the electro-magnetic current for neutrinos takes the
forms
\begin{equation}
\beta=\pi/2  \quad \Longrightarrow \quad
\bar{u}_{\nu_a} (p') i\sigma_{\lambda \rho} q^\rho
\mbox{Re}(\mu_{ea})u_{\nu_e} (p) + \mbox{h.c.}
\label{32}
\end{equation}
\begin{equation}
\beta=0 \quad \Longrightarrow \quad
-\bar{u}_{\nu_a} (p') i\sigma_{\lambda \rho} q^\rho
\mbox{Im}(d_{ea}) \gamma_5 u_{\nu_e} (p) + \mbox{h.c.}
\label{31}
\end{equation}
The first case is the limit that we considered in our previous paper
\cite{asymmetry1}.

\section{Neutrino-electron scattering cross sections}

We consider the scattering of neutrinos on electrons when the initial
flux of neutrinos is not {\em completely polarized}, i.e.~there exists
a mechanism that converts part of the initial left-handed electron
neutrinos (produced in weak processes) into right-handed ones. We
assume that this is a consequence of the presence of non-zero neutrino
electro-magnetic dipole moments. The Sun seems to be the only physical
situation where such depolarization process can occur.

Let us consider the scattering $\nu (k_1) + e^- (p_1) \to \nu
(k_2) + e^- (p_2)$, in the coordinate frame where the
initial electron is at rest.  The four-vectors of the particles
involved, taking into account conservation of momenta, are the
following
$$
k_1=(\omega,\vec{k}_1)\quad
p_1=(m_e,\vec{0})\quad 
k_2=(\omega-T,\vec{k}_2)\quad
p_2=(m_e+T,\vec{p}_2)
$$
where $T$ is the electron recoil energy and $p_1^2=p_2^2=m_e^2$. From
now on we consider the limit of ultra-relativistic neutrinos, i.e.
$k_1^2 = k_2^2 \simeq 0$ and the low-energy limit ($\omega \ll M_W$).
There are two contributions to the scattering process: weak and
electro-magnetic.  As in the previous section we will consider the
case of two neutrino species. There are two inequivalent physical
situations, namely (i) $\nu_e - \nu_\mu$ and (ii) $\nu_e - \nu_s$,
where $\nu_s$ is a sterile type neutrino.

Following the conventions of ref.~\cite{Mandl} for $f=\nu_e,\nu_\mu$,
the corresponding weak matrix amplitudes for Dirac and Majorana
neutrinos are, respectively
$$
M^D_{W_f} = -i2\sqrt{2} G_F \bar{u}^{r'}_f (k_2) \gamma^\mu
\frac{1-\gamma_5}{2} u^r_f (k_1)
\bar{u}^{s'}_e (p_2) \gamma_\mu
\left (g_{fL} \frac{1-\gamma_5}{2} + g_R \frac{1+\gamma_5}{2}\right )
u^s_e (p_1)
$$
\begin{equation}
M^M_{W_f} = i2\sqrt{2} G_F \bar{u}^{r'}_f (k_2) \gamma^\mu
\gamma_5 u^r_f (k_1)
\bar{u}^{s'}_e (p_2) \gamma_\mu
\left (g_{fL} \frac{1-\gamma_5}{2} + g_R \frac{1+\gamma_5}{2}\right )
u^s_e (p_1)
\label{34}
\end{equation}
and obviously zero for sterile neutrinos. Here $g_{eL} = \sin^2
\theta_W + 1/2$, $g_{\mu L} = \sin^2 \theta_W - 1/2$ and $g_R = \sin^2
\theta_W$. The electro-magnetic amplitudes are
\begin{equation}
M^{em}_{ab} = \frac{e}{q^2} \bar{u}^{r'}_b (k_2) \sigma_{\lambda \rho}
q^\rho (\mu_{ab} + id_{ab}\gamma_5) u^r_a (k_1)
\bar{u}^{s'}_e (p_2) \gamma^\lambda u^s_e (p_1)
\label{35}
\end{equation}
where $ab$ denotes $\nu_e\nu_\mu$ or $\nu_e\nu_s$, and the form factors
$\mu_{ab}$ and $d_{ab}$ depend on whether neutrinos are Dirac or
Majorana.

\begin{figure}
\centerline{\protect\hbox{\psfig{file=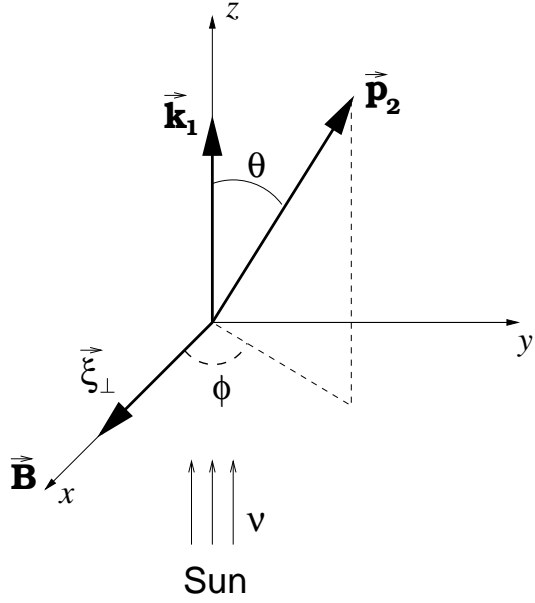,width=7cm}}}
\caption{Coordinate system conventions.}
\label{axis}
\end{figure}

In order to simplify the notation let us set $\mu_a \equiv \mu_{aa}$
and $d_a \equiv d_{aa}$ (both real) for the diagonal case, and $\mu
\equiv \mu_{ea} = \mu^*_{ae}$ and $d \equiv d_{ea} = d^*_{ae}$ for the
transition dipole moments, which are complex in general. We will
perform the calculation of the differential cross section for
neutrino--electron scattering {\sl without} assuming CP invariance and
in the two physical situations. It can be written as a sum of three
terms,
\begin{equation}
\frac{d\sigma}{dTd\phi} = \left
(\frac{d\sigma}{dTd\phi}\right )_{weak} +
 \left (\frac{d\sigma}{dTd\phi}\right )_{em} +
\left (\frac{d\sigma}{dTd\phi}\right )_{int}
\label{totalS}
\end{equation}
that correspond to the purely weak, the purely electro-magnetic and
the interference term, respectively. In the last equation $\phi$ is
the azimuthal angle, as defined in figure \ref{axis}.

\subsection{Active-active case}

In this case ($\nu_e - \nu_\mu$) the purely weak term can be written
in a general form as
\begin{equation}
\label{weak2}
\left (\frac{d \sigma}{d\phi dT}\right )_{weak} = \frac{G_F^2
m_e}{\pi^2} \Bigl [ P_e h(g_{eL},g_R) 
+ P_{\bar{e}} h(g_R,g_{eL})
+ P_\mu h(g_{\mu L},g_R)
+ P_{\bar{\mu}} h(g_R,g_{\mu L})\Bigr ]
\end{equation}
where we have defined $h(x,y) \equiv x^2 + y^2 (1- T/\omega)^2 - xy
m_e T/\omega^2$. Here $P_A$ ($P_{\bar{A}}$) is the probability of
measuring $\nu_A=\nu_{e,\mu}$ ($\bar{\nu}_A=\bar{\nu}_{e,\mu}$), and
from unitarity $P_e + P_{\bar{e}} + P_\mu + P_{\bar{\mu}}=1$.
The purely electro-magnetic term in the presence
of transition dipole moments is
\begin{equation}
\left (\frac{d\sigma}{dTd\phi}\right )_{em} =
\frac{\alpha^2}{2m_e^2}
\left (\frac{1}{T}-\frac{1}{\omega}\right )
\left [\frac{|\mu|^2+|d|^2}{\mu_B^2}
+2 (P_{\bar{e}} + P_\mu - P_e - P_{\bar{\mu}})
\frac{\mbox{Im}(\mu d^*)}{\mu_B^2}
\right ]
\label{transem}
\end{equation}
Note that in the limit of zero mixing only $\nu_e$ and $\bar{\nu}_\mu$
are present ($P_{\bar{e}}=0=P_{\mu}$ and $P_e + P_{\bar{\mu}}=1$), so
that the last equation reduces to the form (see e.~g. \cite{Raffelt2})
\begin{equation}
\left (\frac{d\sigma}{dTd\phi}\right )_{em} =
\frac{\alpha^2}{2m_e^2}
\left (\frac{1}{T}-\frac{1}{\omega}\right )
\frac{|\mu-id|^2}{\mu_B^2}
\label{transem2}
\end{equation}

It will be convenient in order to calculate the interference term of
the cross section to describe the flux of initial neutrinos in terms
of a {\sl density matrix} $\rho$ which generalizes the usual to
account for the case of two different flavors $A,B$. The neutrino part
of the amplitude squared $MM^{'\dagger}$ is thus calculated as follows
\begin{equation}
\sum_{spins} [\bar{u}_B (k_2) M u_A (k_1)]
[\bar{u}_A (k_2) M' u_B (k_1)]^\dagger = \mbox{Tr}[M\rho_{AB} (k_1)
\bar{M}^{'\dagger} \hat{k}_2]
\end{equation}
where $\bar{M}^{'\dagger} = \gamma_0 M^{'\dagger} \gamma_0$ and
$\hat{x} \equiv \gamma_\lambda x^\lambda$.  The density matrix can be
written as a function of the different neutrino probabilities and the
components of the corresponding polarization vectors as follows
\begin{equation}
\rho_{AB} (k_1) = \frac{1}{2} (P_A + P_{\bar{B}} +
(P_{\bar{B}} - P_A)\gamma_5) \hat{k}_1 + 
\frac{1}{4} \hat{\xi}^{\bar{A}B}_\perp \hat{k}_1 (1+\gamma_5) +
\frac{1}{4} \hat{k}_1 \hat{\xi}^{A\bar{B}}_\perp (1-\gamma_5)
\label{39}
\end{equation}
where the polarization four-vectors $\xi^{ \bar{A}B}_\perp=
(0,\vec{\xi}^{\bar{A}B}_\perp)$ are orthogonal to the neutrino
momentum, i.e.  $k_1 \cdot \xi^{\bar{A}B}_\perp = 0 = k_1 \cdot
\xi^{\bar{B}A}_\perp$ and $|\vec{\xi}^{~\bar{A}B}_\perp| =
2\sqrt{P_{\bar{A}}P_B}$.

Note that in the case of one Dirac neutrino $A=B$ this reproduces the
result given in the appendix E of ref.~\cite{Bilenky}, namely
\begin{equation}
\rho_{A} (k_1) = \frac{1}{2} \left( 1 -\xi^{}_\parallel \gamma_5  - 
\hat{\xi}^{}_\perp \gamma_5 \right) \hat{k}_1 
\label{bil}
\end{equation}
where $\vec{\xi}$ is the normalized polarization vector at the
neutrino's rest frame, with $|\vec{\xi^{}_\perp}|=2\sqrt{P_A(1-P_A)}~$
and $\xi^{}_\parallel=2P_A-1$.

It is important to remark that the transversal component
$\vec{\xi}_\perp$ of the neutrino polarization vectors are aligned
along the direction of the solar magnetic field $\vec{B}_\odot$
\cite{Sem97}. 

For ultra-relativistic neutrinos the interference term arises only if
the initial flux contains some mixture of right-handed neutrinos. Note
that in this subsection we restrict to the simple situation where the
initial $\nu_e$ convert to $\bar{\nu}_\mu$ through a neutrino
transition dipole moment (zero neutrino mixing). Then one sets
$P_\mu=P_{\bar{e}}=0$, so that $\vec{\xi}_\perp^{~\bar{e}
\mu}=\vec{0}$. This is the process that can occur in the Sun. The
expression for the interference term is found to be
\begin{eqnarray}
\left (\frac{d\sigma}{dTd\phi}\right )_{int} =
-\frac{\alpha G_F}{4\sqrt{2}\pi m_e T} \Biggl [
\left (\frac{\mbox{Re}(\mu)+\mbox{Im}(d)}{\mu_B}\right )
\vec{p}_2 \cdot \vec{A}_M (T,\omega) \nonumber \\
+ \left (\frac{\mbox{Re}(d)-\mbox{Im}(\mu)}{\mu_B}\right )
\hat{\vec{k}}_1 \cdot 
(\vec{p}_2 \times \vec{A}_M (T,\omega))
\Biggr ]
\label{majtran}
\end{eqnarray}
where $\hat{\vec{k}}_1 \equiv \vec{k}_1/\omega$ and we have defined
\begin{equation}
\vec{A}_M (T,\omega) \equiv \left[(g_{eL} + g_{\mu L} + 2g_R)
\left (2 - \frac{T}{\omega}\right )+
(g_{eL} - g_{\mu L})\frac{T}{\omega }\right ] 
\vec{\xi}_\perp^{e \bar{\mu}}
\label{amaj}
\end{equation}

Note that \eq{majtran} depends explicitly on the azimuthal angle
$\phi$. Choosing the coordinate system as shown in figure \ref{axis},
this dependence is like $\cos \phi$ or $\sin \phi$, since it is easily
checked that
$$
\vec{p}_2\cdot\vec{\xi}_\perp = \mid \vec{p}_2 \mid 
\sin \theta \mid \vec{\xi}_\perp \mid \cos \phi = \sqrt{2m_e
T\left (1 - \frac{T}{T_{max}}\right )} \mid \vec{\xi}_\perp \mid \cos
\phi
$$
\begin{equation}
\hat{\vec{k}}_1 \cdot (\vec{p}_2 \times \vec{\xi}_\perp) = 
-\mid \vec{p}_2 \mid 
\sin \theta \mid \vec{\xi}_\perp \mid \sin \phi = -\sqrt{2m_e
T\left (1 - \frac{T}{T_{max}}\right )} \mid \vec{\xi}_\perp \mid \sin
\phi
\label{angles}
\end{equation}
where $T_{max} = 2\omega^2/(m_e + 2\omega)$ is the maximum electron
recoil energy. 

\subsection{Active-sterile case}

In this case the three terms of the differential cross section in
\eq{totalS} are different with respect to the active-active case,
since sterile neutrinos do not have weak interactions. For instance
the purely weak term will consist only of the electron neutrino
contribution in \eq{weak2}, while the purely electro-magnetic term in
the presence of Dirac-type dipole moments is the well known result
\begin{equation}
\left (\frac{d\sigma}{dTd\phi}\right )_{em} =
\frac{\alpha^2}{2m_e^2}
\left (\frac{1}{T}-\frac{1}{\omega}\right )
\frac{\mu_e^2+d_e^2}{\mu_B^2}
\label{ddem}
\end{equation}
Finally the interference term in the presence of active-sterile dipole
moments is
\begin{eqnarray}
\left (\frac{d\sigma}{dTd\phi}\right )_{int} =
-\frac{\alpha G_F}{2\sqrt{2}\pi m_e T} \Biggl[
\left(\frac{\mbox{Re}(\mu) + \mbox{Im}(d)}{\mu_B}\right )
\vec{p}_2 \cdot \vec{A}_S (T,\omega) \nonumber \\
+ \left(\frac{\mbox{Re}(d) - \mbox{Im}(\mu)}{\mu_B}\right )
\hat{\vec{k}}_1 \cdot (\vec{p}_2 \times \vec{A}_S (T,\omega)) 
\Biggr ]
\label{diractrans}
\end{eqnarray}
where 
\begin{equation}
\vec{A}_S (T,\omega) \equiv \left[g_{eL} + 
g_R\left (1 - \frac{T}{\omega}\right )\right ]
\vec{\xi}_\perp^{~e\bar{s}}
\label{adt}
\end{equation}
In the limit when the active-sterile pair form a Dirac neutrino
\eq{diractrans} reduces to
\begin{equation}
\left (\frac{d\sigma}{dTd\phi}\right )_{int} =
-\frac{\alpha G_F}{2\sqrt{2}\pi m_e T} \left[
\left(\frac{\mu_e}{\mu_B}\right ) \vec{p}_2 \cdot \vec{A}_D (T,\omega)
+ \left(\frac{d_e}{\mu_B}\right ) \hat{\vec{k}}_1 \cdot 
(\vec{p}_2 \times \vec{A}_D (T,\omega))\right ]
\label{diracdiag}
\end{equation}
where
\begin{equation}
\vec{A}_D (T,\omega) \equiv \left[g_{eL} + 
g_R\left (1 - \frac{T}{\omega}\right )\right ]
\vec{\xi}_\perp^{~e}
\label{add}
\end{equation}
The first term in this result was obtained in ref. \cite{Barbieri},
while the second CP-violating term was given in ref.~\cite{SG94} (see
their eq.~9c).

\section{Test of CP conservation at Hellaz}

We propose to measure solar neutrino-electron scattering in upcoming
experiments that will be capable of measuring directionality of the
outgoing $e^{-}$ (like Hellaz). The relevant observable is the
azimuthal distribution of events, namely
\begin{equation}
\label{spectrum}
\frac{dN}{d\phi} = N_e \sum_{i} \Phi_{0i} 
\int^{T_{max}}_{T_{Th}}dT~
\int^{\omega_{max}}_{\omega_{min}(T)}
d \omega~ \lambda_i(\omega) \epsilon(\omega)
\frac{d \sigma}{dTd\phi} (\omega,T) 
\end{equation}
where $d \sigma/dTd\phi$ is the complete differential cross section of
\eq{totalS}, $\epsilon(\omega)$ is the efficiency of the detector
%(which we take as unity for energies above the threshold, for simplicity), 
and $N_e$ is the number of electrons in the fiducial volume of the
detector. The sum in the above equation is done over the solar
neutrino spectrum, where $i$ corresponds to the different reactions
$i= pp$, $^7$Be, $pep$, $^8$B $\ldots$, characterized by a
differential spectrum $\lambda_i(\omega)$ and an integral flux
$\Phi_{0i}$.

In the previous section we found the expressions for the differential
cross section. The azimuthal distribution of the number of events can
be written in a general form as
\begin{equation}
\Frac{dN}{d\phi}=n_{weak}+n_{em}+ n_{int}\cos (\phi +\delta) 
\end{equation}
where $n_{weak}$ ($n_{em}$) accounts for the weak (electro-magnetic)
contributions, while $n_{int}$ is the interference term. The dependence
of the last term on the azimuthal angle $\phi$ is parametrized with
$\delta$. Thus a pure $\cos \phi$ ($-\sin \phi$) dependence corresponds
to $\delta=0$ ($\delta=\pi/2$). We can define the
differential azimuthal asymmetry as
\begin{equation}
\left.\Frac{dA}{d\phi}\right|_{\phi '}=
\Frac{\left.\Frac{dN}{d\phi}\right|_{\phi '}-
\left.\Frac{dN}{d\phi}\right|_{\phi '+\pi}}{
\left.\Frac{dN}{d\phi}\right|_{\phi '}+
\left.\Frac{dN}{d\phi}\right|_{\phi '+\pi}}=
\Frac{n_{int}}{n_{weak}+n_{em}} \cos (\phi ' + \delta)
\end{equation}
where $\phi$ ($\phi '$) is measured with respect to the direction of
the magnetic field $\vec{B}_\odot$, which we will assume to be along
the positive $x$-axis (see fig.~\ref{axis}). By integrating over
$\phi$ one can also define an asymmetry $\cal A$ as
\begin{equation}
{\cal A}(\phi ')=
\Frac{{\displaystyle \int^{\phi '+\pi}_{\phi '}} \Frac{
dN}{d\phi}d\phi - {\displaystyle \int^{\phi '+2\pi}_{\phi '+\pi}}
\Frac{ dN}{d\phi}d\phi}
{{\displaystyle \int^{\phi '+\pi}_{\phi '}} \Frac{
dN}{d\phi}d\phi + {\displaystyle \int^{\phi '+2\pi}_{\phi '+\pi}}
\Frac{ dN}{d\phi}d\phi}=
-A \sin (\phi ' + \delta)
\label{asym4}
\end{equation}
where $A\equiv 2 n_{int}/\pi (n_{weak}+n_{em})$ is the maximum
integrated asymmetry measurable by the experiment, which is manifestly
positive. In our previous paper \cite{asymmetry1} we calculated the
expected values of $A$ for $pp$ solar neutrinos at Hellaz, for
different choices of the survival probability of $\nu_e$'s, in the
CP-conserving case of \eq{32}.

Let now discuss how the measurement of the azimuthal asymmetry could
be carried out considering that $\vec{B}_\odot$ is constant over a
given period of time {\em and} its direction is known. One should
collect events in every $\phi$-bin, where $\phi$ is defined with
respect to the positive $x$-axis and then take for different $\phi 's$
the ratio ${\cal A} (\phi)$ which should show a $\sin \phi$ dependence
with a maximum equal to $A$.  This will allow us to identify the value
of $\delta$.

Note that if we were able to find from the measurements that $\delta
\neq 0$ beyond experimental uncertainties, then this would lead to the
conclusion that {\sl CP is not conserved in the electro-magnetic
interactions of neutrinos} if we consider Dirac diagonal or Majorana
transition dipole moments. In the Dirac transition case the CP phases
$\eta_i$ of the neutrinos can be chosen so as to have CP conservation
for any value of $\delta$. However, Dirac dipole moments do not seem
to be favored by theoretical models nor by the existent astrophysical
and cosmological constraints \cite{Raffelt}.

\section{Discussion}

We have shown that the scattering of solar neutrinos on electrons may
reveal their CP properties, which are particularly sensitive to their
Majorana nature, due to the interference of electro-magnetic and weak
amplitudes.  We showed  how future solar neutrino experiments with good
angular resolution and low energy threshold can be sensitive to the
resulting azimuthal asymmetries in event number, and could therefore
provide valuable information on the CP properties and the nature of
the neutrinos, provided the solar magnetic field direction is fixed.
Hellaz will be the first experiment which is potentially sensitive to
azimuthal asymmetries since the directionality of the outgoing $e^{-}$
can be measured. The angular resolution is expected to be $\Delta
\theta \sim \Delta \phi \sim 30$ mrad $\sim 2^\circ$, substantially
better than that of Super-Kamiokande.  Notice also that the width of
the Cerenkov cone defined by the angle $\theta$ is very narrow for
high-energy boron neutrinos, as one can see from \eq{angles}.  In
contrast, for $pp$ neutrino energies accessible at Hellaz ($T_{max}
\simeq 0.26$ MeV, $T_{th} \simeq 0.1$ MeV) we estimate that $\theta$
can be as large as $48^\circ$.
It is important to emphasize here that, while the existence of an
asymmetry in event number can be ascribed to a non-zero neutrino
electro-magnetic dipole moment, one can not infer any information on
the specific issue of CP conservation in the neutrino sector and the
nature of neutrinos without an accurate knowledge of the direction of
the solar magnetic field. Such a knowledge is indeed possible except
at minimal solar activity periods, when the toroidal magnetic field
vanishes.

\section*{Acknowledgements}

The authors thank Thomas Ypsilantis for fruitful discussions on the
Hellaz experiment.  This work has been supported by DGICYT under
Grants PB95-1077, by the TMR network grant ERBFMRXCT960090 and by
INTAS grant 96-0659 of the European Union. V.~S. acknowledges also the
support of Generalitat Valenciana and RFBR grant 97-02-16501.

\end{document}